\documentclass{appolb}
\usepackage{xcolor}
\usepackage{amsmath}
\usepackage{epsfig}

\usepackage{textpos}

\begin{document}

\title{Exploring the In--Medium Momentum Dependence of the Dynamical Quark Mass
}
\author{Mateusz Cierniak
\and
Thomas Kl\"{a}hn
\address{Division of Elementary Particle Theory\\
Institute of Theoretical Physics\\
University of Wroc\l aw}
}
\maketitle

\begin{abstract}
We review the two standard equations of states based on the Nambu--Jona-Lasinio (NJL) model and the thermodynamic bag (tdBag) model  for dense, cold quark matter from a perspective based on the Dyson--Schwinger (DS) formalism.
A different, but technically not more complicated approximation reproduces the  model of Munczek and Nemirovsky (MN) which 
accounts in a simplified way for chiral symmetry breaking and confinement as a dynamic process
rooted in the momentum dependence of QCD model gap solutions. We review the mass gap solutions for the MN model in the chiral limit
and sketch the behavior of mass gap solutions for finite bare quark masses at finite chemical potential.
\end{abstract}

\section{Introduction}

It is believed that QCD is the correct theory of strongly interacting matter. 
Key properties of QCD that need to be addressed in a realistic model are confinement and chiral symmetry breaking. 
Both of these effects are an important aspect of the strong interaction which is probed in heavy ion collisions and
increasingly so by astrophysical observations. 
The latter provide an interesting alternative to test our knowledge regarding the equation of state of nuclear matter
and the expected QCD phase transition to a quark gluon plasma \cite{Oertel:2016bki}.
Lattice QCD calculations provide insight into the QCD phase space at high temperatures in vacuum or at low baryochemical potential. 
They fail when the quark chemical potential exceeds the temperature by far, as it is given in compact stars. 
The Dyson--Schwinger approach to QCD is applicable in the entire temperature--density domain.
It requires an appropriate truncation scheme, since the explicit set of Dyson--Schwinger equations is infinite and thus in the strict sense unsolvable. 
In this brief review we emphasize the fact, that the NJL model \cite{Nambu_61a, Nambu_61b}
as one of the state of the art equations of state 
can be easily understood in terms of DS equations within a set of very simple approximations.
The price for the convenient description of chiral symmetry breaking is paid for with the absence
of any momentum dependence of the DS gap functions which reflects the well known fact that
the NJL model does not exhibit confinement.
Once chiral symmetry is restored the NJL mass gap solution provides a nearly constant quark mass
which is much smaller than the quark chemical potential in this domain.
Consequently the equation of state is well approximated by an ideal relativistic Fermi gas shifted
by a constant offset with respect to the pressure and energy density respectively \cite{Klahn:2015mfa,Klahn:2016uce}.
Formally, this corresponds to the behavior described by the thermodynamic bag model \cite{Farhi:1984qu}.
None of these two effective models has mass gap solutions with a nontrivial momentum dependence, 
viz. solutions other than constant or zero for any momentum at any given density.
Consequently, within these models a confinement criterion that implies the absence of quark mass poles is 
impossible to account for  and the deconfinement 
transition has to be modeled by imposing additional assumptions.
We review properties of the similary simple but confining MN model in the chiral limit \cite{Munczek:1983dx} and explore 
mass gap solutions at finite bare quark masses and finite chemical potential.
Computations like these are a prerequisite to study the quark matter EoS within the MN model beyond those currently available in
the chiral limit \cite{Blaschke:1997bj, Klahn:2009mb}.

\section{Dyson--Schwinger Equations}

The in-medium, dressed-quark propagator maintains the structure of
a free, relativistic Fermion propagator,
\begin{equation}
S(p^2, \tilde p_4)^{-1}=i\vec{\gamma}\vec{p}A(p^2, \tilde p_4)+i\gamma_{4}\tilde{p}_{4} C(p^2, \tilde p_4)+B(p^2, \tilde p_4),
\end{equation} 
with $\tilde p_4=p_4+i\mu$.
Evidently, the gap functions $A$, $B$, and $C$ account for non-ideal behaviour due to interactions. 
Unlike in vacuum studies, the gaps are complex valued and $A$- and $C$-gap are degenerate ($A$=$C$ holds strictly under vacuum conditions).
In order to obtain the propagator one solves the gap equation
\begin{eqnarray}
S(p^2,\tilde p_4)^{-1} & = & i \vec{\gamma}\cdot \vec{p} + i\gamma_4 \tilde p_4 +m + \Sigma(p^2,\tilde p_4)\,,\\
\nonumber
\Sigma(p^2, \tilde p_4 ) &=& \int\frac{d^4 q}{(2\pi)^4}\, g^2(\mu) D_{\rho\sigma}(p-q,\mu) 
\frac{\lambda^a}{2}\gamma_\rho S(q^2,\tilde q_4) \Gamma^a_\sigma(q,p,\mu) , \label{gensigma}
\end{eqnarray}
where $m$ is the bare mass, $D_{\rho\sigma}(k,\mu)$ is the dressed-gluon propagator and $\Gamma^a_\sigma(q,p,\mu)$ is the dressed-quark-gluon vertex. 
Naturally, at the level of the self energy $\Sigma(p^2, \tilde p_4 )$ approximations can be made in order to
simplify the gap equations.
In all following discussions we impose 
\begin{equation} 
\Gamma^a_\sigma(q,p)=\frac{1}{2}\lambda^a\gamma_\sigma
\end{equation}  
for the vertex and thus define a rainbow gap equation, 
which is the leading-order in a systematic, 
symmetry-preserving DSE truncation scheme \cite{Munczek:1994zz,Bender:1996bb}.

We can now introduce NJL and MN model in terms of different choices for the effective gluon-propagator.

\section{NJL and tdBag model}
The NJL model can be strictly understood in terms of a contact interaction in configuration space 
provided by the gluon propagator. Transformed into momentum space this reads as a constant,
momentum-independent coupling. As a consequence this model is ultraviolet-divergent
if no regularization is performed. In the spirit of the standard NJL approach we perform a hard cut-off
in the UV and express the effective gluon propagator as
\begin{equation}
g^{2}D^{\rho\sigma}(p-q)=\frac{1}{m^{2}_{G}}\Theta(\Lambda^{2}-\vec{q}^{2})\delta^{\rho\sigma}.
\end{equation}
The Heaviside function $\Theta$ provides a 3-momentum cutoff for space-like momenta $\vec p^2>\Lambda^2$.
This is sufficient to regularize all ultraviolet divergences inherent to $\Sigma(p^2, \tilde p_4 )$.
Different regularisation procedures are available and in fact the regularisation scheme  
does not have to affect ultraviolet divergencies only. 
E.g., IR cutoff schemes can remove unphysical implications \cite{Ebert:1996vx}.
However, the chosen hard cut-off scheme
reproduces standard NJL model results and allows to match them to tdBAG, 
i.e. to describe quarks as a quasi ideal gas of Fermions.
$m_G$ is a gluon mass scale which in this model simply defines the coupling strength. 
These approximations are sufficient to write the gap equations.
For the $A$-gap follows the trivial, medium independent solution, $A=1$. 
The remaining gap equations take the following form,
\begin{eqnarray}
\label{eq:BGap}
B_p &=& m + \frac{16N_c}{9m_G^2}\int_\Lambda\frac{d^4q}{(2\pi)^4}
\frac{ B_q}{\vec q^2A^2_q+ \widetilde q^2_4C^2_q + B^2_q}~,
\\
\label{eq:CGap}
\widetilde p_4^2 C_p &=& \widetilde p_4^2 + \frac{8N_c}{9m_G^2}\int_\Lambda\frac{d^4q}{(2\pi)^4}
\frac{\widetilde p_4\widetilde q_4C_q}{\vec q^2A^2_q+ \widetilde q^2_4C^2_q + B^2_q}~.
\end{eqnarray}
The integrals do not explicitly depend on the external momentum $p$ and consequently, both gap solutions are constant at any given $\mu$. 
Both equations can be recasted
\begin{eqnarray}
\label{EQ:massgapscalar}
B &=& m+\frac{4N_c}{9m_G^2}n_s(\mu^*,B)\\
\label{EQ:mugapvector}
\mu &=& \mu^*+\frac{2N_c}{9m_G^2}n_v(\mu^*,B),
\label{eq:gaps_DS}
\end{eqnarray}
in terms of the single-flavor scalar and vector densities, $n_s$ and $n_v$, of an ideal spin-degenerate Fermi gas,
\begin{eqnarray}
n_s &=& 2\sum_\pm\int_\Lambda\frac{{\rm d}^3 \vec p}{(2\pi)^3}\frac{B}{E}
\left(
\frac{1}{2}-\frac{1}{1+\exp{(E^\pm/T)}}
\right), \\
n_v &=& 2\sum_\pm\int_\Lambda\frac{{\rm d}^3 \vec p}{(2\pi)^3}\frac{\mp1}{1+\exp{(E^\pm/T)}},
\end{eqnarray}
with $E^2=\vec p^2+B^2$ and $E^\pm=E\pm\mu^*$.
The merit of the NJL model is the ability to describe chiral symmetry breaking 
as the formation of a scalar condensate, and chiral symmetry restoration as the melting of
the same. It should be kept in mind though, that it is the scalar density which requires
UV regularization and in that sense chiral symmetry breaking can be considered as
the most sensitive part of the model.

The next information the latter equations provide is not new. 
The NJL model describes quarks as quasi-ideal particles with corresponding quasi particle poles.
Confinement is not accounted for. 
In \cite{Klahn:2015mfa} it has been pointed out how this can be understood as a reason for 
the fact that NJL models typically provide a larger bag constant than, e.g., the MIT-bag model
would require. The NJL model does not 'bind'. Adding, or better substracting the missing 
binding energy per volume to the equation of state can lead
to interesting results over the whole phase diagram, as illustrated in \cite{Klahn:2016uce}.

\section{MN Model}
The underlying approximation for this model is a gluon propagator with constant 
strength over the whole configuration space. 
The momentum-dependent Fourier-transform of this object therefore reads as
\begin{equation}
g^{2}D^{\rho\sigma}(k)=3\pi^{4}\eta^{2}\delta^{\rho\sigma}\delta^{(4)}(k),
\end{equation}
with $\eta$ representing the strength of the effective interaction. 
The ansatz was proposed in \cite{Munczek:1983dx} and extended for non-zero chemical potential in \cite{Klahn:2009mb}. In both cases, the considerations were limited to chiral quarks. 
Although the assumption of a model with support only at $p=0$, hence {\it only} infrared strength,
is certainly peculiar
it results in a model with interesting features: i) it is UV-finite and does not require regularisation
or renormalisation and it has ii) momentum dependent gaps which give access to distinct phases one can 
interpret as confined/chirally broken and deconfined/chirally restored.

In the chiral limit, the nonperturbative, chiral symmetry preserving solution of the gap equation is $\hat A(p^2,\tilde p_4)= \hat C(p^2,\tilde p_4)$,
\begin{equation}
\hat C(p^2,\tilde p_4) =\frac{1}{2}
\left(
	1+\sqrt{1+\frac{2\eta^2}{\tilde p^2}}
\right)\,,\; \hat B(p^2,\tilde p_4)  \equiv  0\,,
\label{CABW}
\end{equation}
where $\tilde p^2 = \vec p^2 + (p_4+i\mu)^2$.  Here, chiral symmetry is realised in the Wigner-Weyl mode and the quark is not confined.

The gap equation also has a confining solution with dynamically broken chiral symmetry for $m=0$,
\begin{eqnarray}
\label{CNG}
C(p^2,p\cdot u) & = &
\left\{
\begin{array}{ll}
2 & \rm{Re}(\tilde p^2) < \frac{\eta^2}{4}\\
\frac{1}{2}
\left(
	1+\sqrt{1+\frac{2\eta^2}{\tilde p^2}}
\right) & \mbox{otherwise,}
\end{array}
\right.\\
\label{BNG}
B(p^2,p\cdot u) &=&
\left\{
\begin{array}{ll}
\sqrt{\eta^2 - 4 \tilde p^2} & \rm{Re}(\tilde p^2) < \frac{\eta^2}{4}\\
0 & \mbox{otherwise.}
\end{array}
\right.
\end{eqnarray}
Here, chiral symmetry is realised in the Nambu-Goldstone mode.  Confinement is signalled by a square-root branch point at $\tilde p^2 = \eta^2/4$, associated with the scalar piece of the self energy.  For $\mu \neq 0$, it occurs at $p_4=0$, $\vec{p}\,^2=\mu^2+\eta^2/4$.

\section{MN mass gaps for non-chiral bare quarks}

\begin{figure}[h]
\centering
\includegraphics[width=0.9\linewidth]{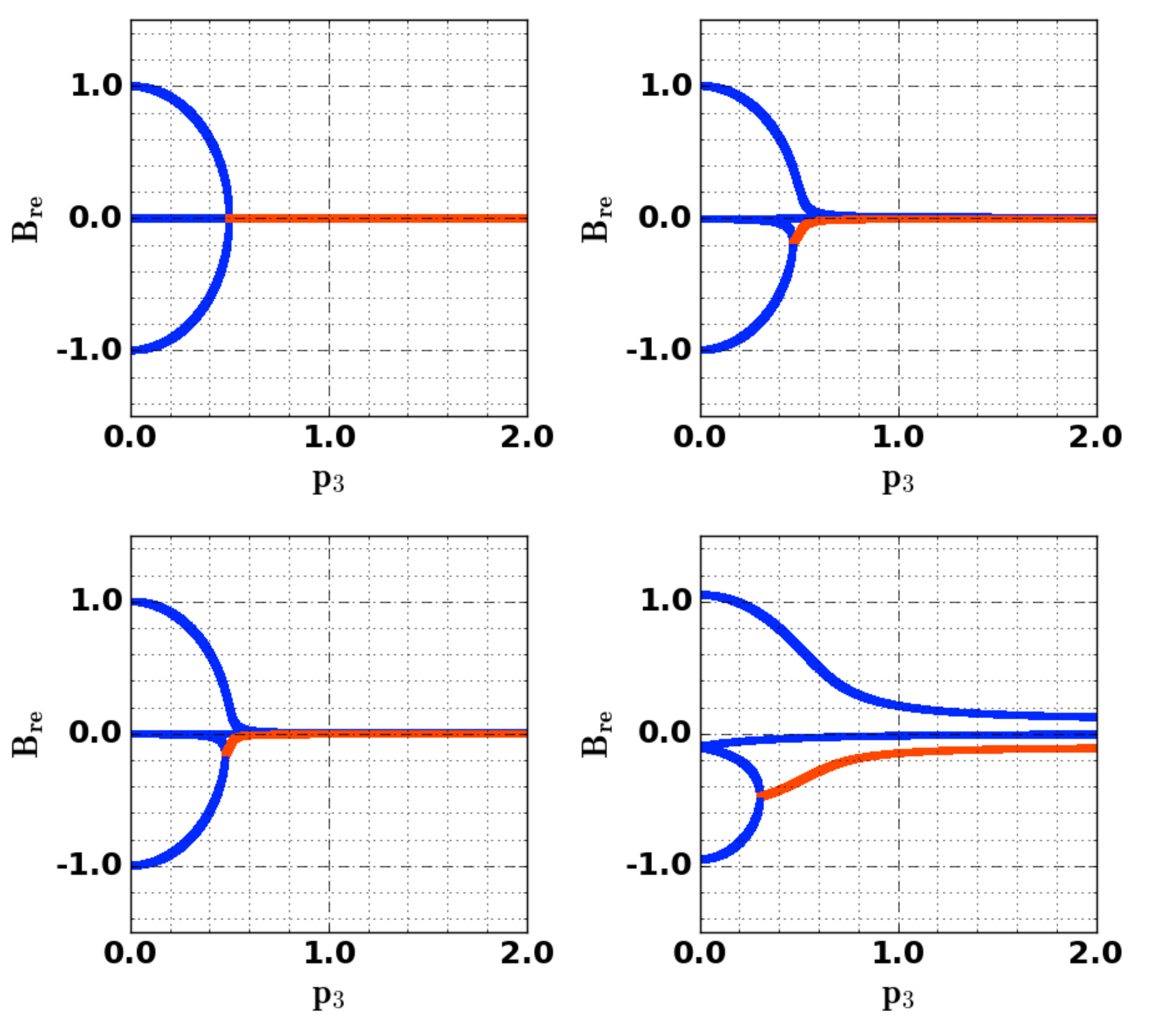}
\caption[The solution of MN gap equations as a function of momentum for various bare masses.]{The solution of MN gap equations as a function of momentum (with $\mu=p_{4}=0$). Blue color represents real solutions and red complex. Top left - chiral quark($m=0$), top right - up quark ($m=3$ MeV), bottom left - down quark ($m=5$ MeV), bottom right - strange quark ($m=100$ MeV). All values are in units of $\left[\frac{GeV}{\eta}\right]$.}
\label{fig:mn_res1}
\end{figure}

Imposing an explicit finite bare mass term, the mass gap equation can be fully expressed in terms of the free variables, taking the polynomial form
\begin{equation}\label{blabla2}
B^{4}+mB^{3}+B^{2}(4\tilde{p}^{2}-m^{2}-\eta^{2})-mB(4\tilde{p}^{2}+m^{2}+2\eta^{2})-\eta^{2}m^{2}=0.
\end{equation}
Note, that this prescription differs from the original vacuum result provided in  \cite{Munczek:1994zz} only by the appearance of the chemical potential,
$p_4\to\tilde p_4= p_4 + i\mu$ in $p^2=\vec p^2+p_4^2$.
Evidently, this is sufficient to generate complex mass gap solutions.
The four solutions of this polynomial equation are the possible quark effective masses. Their momentum dependence in vacuum is shown in Fig.\eqref{fig:mn_res1}
The top left figure shows good agreement with analytical chiral limit results in \cite{Klahn:2009mb} with a clear discontinuous transition from a massive to a massless branch. The addition of non--zero bare mass changes the qualitative behavior of the solutions, as the high--low mass transition is now smooth. Furthermore, one of the chiral solution appears to be degenerate. 
This degeneracy is lost for a finite bare mass. Despite this, small bare mass solutions show approximate agreement with the chiral solutions, especially for the positive branch. This illustrates the impact of dynamic chiral symmetry breaking on the effective mass of massive quarks and justifies the approximation of light quarks as massless, at the same time showing that such an approximation is increasingly questionable for quarks with masses of the order of $0.1$ GeV and above.

\begin{figure}[h]
\centering
\includegraphics[width=0.45\linewidth]{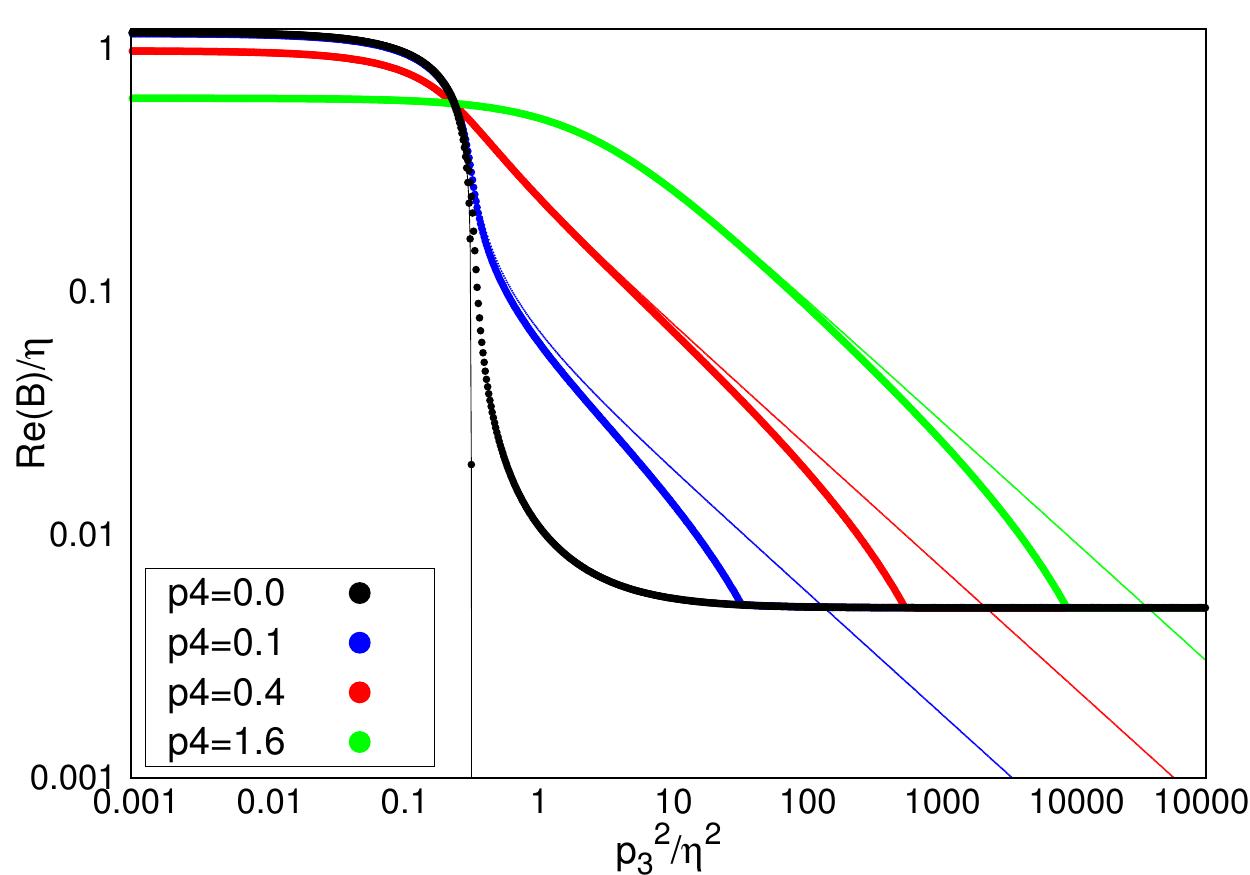}
\includegraphics[width=0.45\linewidth]{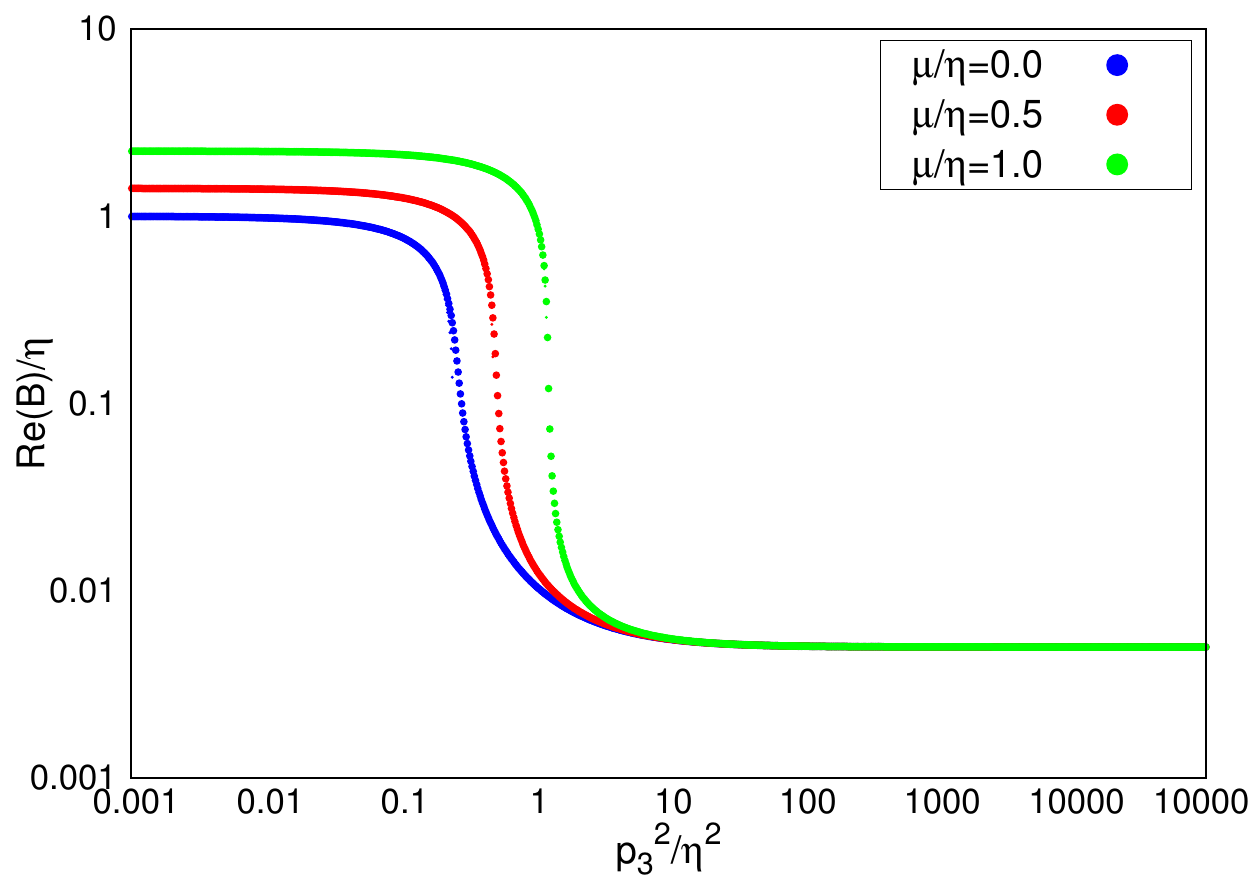}
\caption[The solution of MN gap equations as a function of momentum for various energies and chemical potentials.]{The solution of MN gap equations as a function of momentum for varying $p_{4}$ and $\mu$. Thin lines - $m=0$, bold lines - $m=5$ MeV. All values are in units of $[GeV]$.}
\label{fig:mn_res2}
\end{figure}

The effective mass is sensitive to both, energy and chemical potential, as illustrated in Fig.\eqref{fig:mn_res2} for the positive mass branch.
The non--zero mass solutions exhibit a sharp transition at high 3-momentum and finite energy. 
This transition is not observed in the chiral limit or in the case of zero energy. 
The effect of increasing the chemical potential is an increased value of the mass gap value at all momenta.

\section{Conclusions}

The results presented in this work show the remarkable utility of the Dyson--Schwinger equations in deriving in--medium properties of a theory, a task notoriously difficult using lattice methods. 
Two models of a gluon propagator were used, NJL (or bag-like) models with constant interaction strength in momentum space and the MN model with infrared strength only. 
The former was used as a proof--of--concept test for the Dyson--Schwinger formalism and has shown good agreement with existing effective models. The latter, an extension of the model proposed by \cite{Munczek:1983dx} gave the opportunity to study quark properties in--medium. The results have shown good agreement with previous studies of this model \cite{Klahn:2009mb}. 
Further, the model has a rich structure when combining non--zero bare quark mass, finite energy and chemical potential. The results underline the importance of infrared interactions on the properties of strongly interacting matter and warrant a more in--depth study of this models possible extensions.

\section{Acknowledgements}
MC and TK acknowledge support from the Polish  National  Science  Center  (NCN)  under  grant
numbers UMO-2013/09/B/ST2/01560  and UMO-2014/13/B/ST9/02621  respectively. The authors acknowledge support from the COST Action MP1304 ”NewCompStar” for their networking activities.

\newpage

\bibliography{main_tk.bbl}

\begin{thebibliography}{10}

\bibitem{Oertel:2016bki}
M.~Oertel, M.~Hempel, T.~Klähn, and S.~Typel.
\newblock {Equations of state for supernovae and compact stars}.
\newblock 2016.

\bibitem{Nambu_61a}
Yoichiro Nambu and G.~Jona-Lasinio.
\newblock {DYNAMICAL MODEL OF ELEMENTARY PARTICLES BASED ON AN ANALOGY WITH
  SUPERCONDUCTIVITY. II}.
\newblock {\em Phys.Rev.}, 124:246--254, 1961.

\bibitem{Nambu_61b}
Yoichiro Nambu and G.~Jona-Lasinio.
\newblock {Dynamical Model of Elementary Particles Based on an Analogy with
  Superconductivity. I.}
\newblock {\em Phys.Rev.}, 122:345--358, 1961.

\bibitem{Klahn:2015mfa}
Thomas Klahn and Tobias Fischer.
\newblock {Vector interaction enhanced bag model for astrophysical
  applications}.
\newblock {\em Astrophys. J.}, 810(2):134, 2015.

\bibitem{Klahn:2016uce}
Thomas Klahn, Tobias Fischer, and Matthias Hempel.
\newblock {Simultaneous chiral symmetry restoration and deconfinement -
  Consequences for the QCD phase diagram}.
\newblock {\em Astrophys. J.}, 836(1):89, 2017.

\bibitem{Farhi:1984qu}
Edward Farhi and R.~L. Jaffe.
\newblock {Strange Matter}.
\newblock {\em Phys. Rev.}, D30:2379, 1984.

\bibitem{Munczek:1983dx}
H.~J. Munczek and A.~M. Nemirovsky.
\newblock {The Ground State q anti-q Mass Spectrum in QCD}.
\newblock {\em Phys. Rev.}, D28:181, 1983.

\bibitem{Blaschke:1997bj}
David Blaschke, Craig~D. Roberts, and Sebastian~M. Schmidt.
\newblock {Thermodynamic properties of a simple, confining model}.
\newblock {\em Phys. Lett.}, B425:232--238, 1998.

\bibitem{Klahn:2009mb}
Thomas Klahn, Craig~D. Roberts, Lei Chang, Huan Chen, and Yu-Xin Liu.
\newblock {Cold quarks in medium: an equation of state}.
\newblock {\em Phys. Rev.}, C82:035801, 2010.

\bibitem{Munczek:1994zz}
H.~J. Munczek.
\newblock {Dynamical chiral symmetry breaking, Goldstone's theorem and the
  consistency of the Schwinger-Dyson and Bethe-Salpeter Equations}.
\newblock {\em Phys. Rev.}, D52:4736--4740, 1995.

\bibitem{Bender:1996bb}
A.~Bender, Craig~D. Roberts, and L.~Von~Smekal.
\newblock {Goldstone theorem and diquark confinement beyond rainbow ladder
  approximation}.
\newblock {\em Phys. Lett.}, B380:7--12, 1996.

\bibitem{Ebert:1996vx}
Dietmar Ebert, Thorsten Feldmann, and Hugo Reinhardt.
\newblock {Extended NJL model for light and heavy mesons without q - anti-q
  thresholds}.
\newblock {\em Phys. Lett.}, B388:154--160, 1996.

\end{thebibliography}

\end{document}